%% file: paper.tex
\documentclass[aip,pop,reprint]{revtex4-1}

\usepackage{amssymb}
\usepackage{array}
\usepackage{xcolor}
\usepackage{graphicx}
\usepackage{bm}
\usepackage{subcaption}
\usepackage{afterpage}
\usepackage{amsmath}
\usepackage{float}

\newcommand\angfreq[1]{\omega_\text{#1}}
\newcommand\period[1]{\tau_\text{#1}}
\newcommand\mass[1]{m_\text{#1}}
\newcommand\charge[1]{q_\text{#1}}
\newcommand\density[1]{n_\text{#1}}
\newcommand\velocity[1]{v_\text{#1}}
\newcommand\relvelocity[1]{\beta_\text{#1}}
\newcommand\temperature[1]{T_\text{#1}}
\newcommand\kb{k_\text{B}}
\newcommand\debye{\lambda_\text{D}}

\input{preamble.tex}
\usepackage[capitalize]{cleveref}

\begin{document}
\graphicspath{ figures }
%\preprint{AIP/123-QED}

\title{Three-Dimensional Particle-In-Cell Simulations of Two-Dimensional Bernstein-Greene-Kruskal Modes}
% Force line breaks with \\
\author{M.
    T.
    Franciscovich}\email[]{mtfranciscovich@alaska.edu}
    \affiliation{Geophysical
          Institute, University of Alaska Fairbanks,
          Fairbanks, Alaska 99775, USA}

    \author{J.
    McClung}\email[]{james.mcclung@unh.edu}
      \affiliation{Department of Physics, University of
          New Hampshire}

      \author{K.
    Germaschewski}\email[]{kai.germaschewski@unh.edu}
    \affiliation{Department of
          Physics, University of New Hampshire}

      \author{C.
    S.
    Ng}\email[]{cng2@alaska.edu}
%\homepage[]{https://sites.google.com/a/alaska.edu/chungsangng/}
\affiliation{Geophysical Institute, University of Alaska Fairbanks, Fairbanks, Alaska 99775, USA}

\date{\today}

\begin{abstract}
In this paper, we present three-dimensional (3D) Particle-In-Cell (PIC) simulations to study the stability of 2D Bernstein-Greene-Kruskal (BGK) modes in a magnetized plasma with a finite background magnetic field. The simulations were performed using the Plasma Simulation Code (PSC) [Germaschewski et al., J. of Comp. Phys. 318, 305 (2016)], as in our recent study using 2D PIC simulations [McClung et al., Phys. Plasmas 31, 042302 (2024)], in order to see if and how the previous results would change with 3D effects. We found that solutions that are stable (unstable) in 2D simulations are still stable (unstable) in the new 3D simulations. However, the instability develops slower in 3D than in 2D and forms an unstable spiral wave structure that is in-phase along the axial direction. We have also simulated cases with an electron density bump (EDB) at the center, in addition to cases with an electron density hole (EDH) considered in our previous study, and found differences in the unstable spiral wave structures between the two cases. Additionally, we have generalized our simulations to have an increased electron thermal velocity, as well as using initial conditions solved from the complete Vlasov-Maxwell system of equations. We found that these generalizations did not change the overall behavior of the simulations and the instability that evolves. 

\end{abstract}

\maketitle

\section{\label{sec:intro}Introduction\protect}

High-temperature plasmas relevant to fusion experiments, space physics, and
astrophysics can be considered collisionless due to small collision frequency. \cite{RevModPhys.71.S404} In these high-temperature, collisionless plasmas, small-scale kinetic structures known as Bernstein-Greene-Kruskal (BGK) modes can exist.\cite{PhysRev.108.546} These structures are described theoretically as nonlinear exact steady-state solutions to the Vlasov-Poisson system of equations. When BGK first explored these solutions, they were limited to one dimension, with the solution being spatially variant along a single Cartesian coordinate, and the solutions only involved solving the Vlasov-Poisson system of equations. Furthermore, these solutions relied on electron trapping as a mechanism for the existence of such structures. Later, these analytic solutions were generalized in three dimensions (3D) for an unmagnetized plasma with a spherically symmetric electric potential localized in all spatial directions.\cite{PhysRevLett.95.245004} Another analytic solution of BGK modes was solved in two dimensions (2D) using a magnetized plasma that had a uniform background magnetic field with a cylindrically symmetric electric potential that was localized on a 2D plane perpendicular to the background magnetic field.\cite{doi:10.1063/1.2186187,
      doi:10.1063/1.5126705} The list of research on BGK modes is extensive. Please see Ref.~\onlinecite{doi:10.1063/1.5126705} for more literature in this area of research.

In a recent paper by McClung et al.,\cite{10.1063/5.0187853} more work was done on simulating the stability of the 2D BGK mode. In that paper, it was shown that the stability of such solutions relied on the strength of the background magnetic field. As the background magnetic field strength increased, the BGK mode became more stable. These high-resolution simulations were conducted using the fully electromagnetic Particle-In-Cell (PIC) code Plasma Simulation Code (PSC) running in parallel on supercomputers utilizing a grid of up to $2048^2$ cells. \cite{GERMASCHEWSKI2016305} A natural and necessary extension of these simulations and results in 2D is to simulate these solutions with an additional third dimension along the axial direction of the background magnetic field to study possible 3D effects. 

In this paper, we present the 3D simulations for the same set of parameters that created the ``electron density hole" (EDH) used in McClung et al.\cite{10.1063/5.0187853} We will also present several generalizations we have made to the solution and simulations. First, we have included an additional set of parameters for the solution that leads to a separate case for the initialized structure in the simulation. This set of parameters results in an ``electron density bump" (EDB) initially instead of the EDH seen in our previous paper. Next, we have increased the electron thermal temperature in the simulations by a factor of ten. Finally, instead of only solving the Vlasov-Poisson (VP) system of equations to find the form of the initial conditions for the simulation, we have generalized the simulation code to include a non-uniform background magnetic field that is solved from the entire Vlasov-Maxwell (VM) system of equations, which includes Poisson's equation and Amp\`ere's law. 

The main objectives of this research are to confirm the validity of analytic solutions, quantify how the stability of such modes depends on key parameters, such as the background magnetic field strength, and to observe the dynamical evolutions of the modes after they become unstable as a result of the added simulation dimension. We also wanted to see if our generalizations to both the solution and the simulation impact the nature of the instability.

We found that the stability of the structure still directly depends on the strength of the background magnetic field, even with the inclusion of the third simulation dimension. For the EDH, our simulations show that the instabilities and azimuthal electrostatic waves seen in 2D are also present in 3D; however, the delay of the onset of the instability and the in-phase nature of the instability at all locations along the axial direction require explanation. Also, we found that the EDB simulations produced a somewhat different instability pattern than the one seen in the EDH case. In particular, we found that the EDB results in a more complicated spiral instability, than the one seen with the EDH.\cite{10.1063/5.0187853}  The simulations showed that our generalization of increased electron thermal velocity had no impact on stability. Lastly, the inclusion of solving Amp\`ere's law only changed the time-steadiness of certain field quantities, and had no impact on overall stability.

In Sec. II, a summary of the 2D BGK mode will be presented. In Sec. III, the setup of the simulations will be discussed, and more thorough definitions of our simulation cases will be given. In Sec. IV, we will show and discuss our results. Finally, we will present our conclusions in Sec. V.

\section{\label{sec:BGK_SOLUTION}BGK MODE SOLUTION\protect}

\subsection{\label{sec:Theory}Theory}

The collisionless Boltzmann equation, or the Vlasov equation, is an equation that describes the evolution of a distribution function $f$ in an $N$-dimensional phase space, where $N$ is the number of spatial and momentum coordinates being considered. Using vector quantities for velocity, \textbf{v}, position, \textbf{r}, and acceleration, \textbf{a}, it reads:
\begin{equation}
    \frac{\partial f}{\partial t} + \textbf{v} \cdot \nabla_r f + \textbf{a} \cdot \nabla_v f = 0. 
\end{equation}
BGK modes are exact, nonlinear solutions to the time-independent form of this equation where the acceleration is given by the Lorentz force equation so that it becomes, for species $s$,
\begin{equation}
    \textbf{v} \cdot \nabla_r f_s + \frac{q_s}{m_s}(\textbf{E} + \textbf{v}\times\textbf{B}) \cdot \nabla_v f_s = 0, 
\end{equation}
where $f_s$ is the distribution function describing the electron species ($s = e$) or the ion species ($s = i$), and $q_s$ and $m_s$ are the associated charge and mass for each species, respectively. In this paper, we utilize only a single ion species. Solutions to this equation coupled with Poisson's equation,
\begin{equation}
    \nabla^2\Psi = -\frac{\rho_q}{\epsilon_0},
\end{equation}
for a charge density $\rho_q$, are what is typically needed for BGK mode solutions. As mentioned previously, we will also make use of the time-steady Amp\`ere's law for initialization, given here as:
\begin{equation}
    \nabla\times\textbf{B}=\mu_0\textbf{J}
\end{equation}
Using the zeroth-order moment of the distribution function, $\rho_q = \sum_s q_s \iiint f_s(\textbf{r},\textbf{v}) \,d^3v$, Poisson's equation becomes:
\begin{gather}
      \nabla^2\Psi = -\frac{1}{\epsilon_0}\sum_s q_s \iiint f_s(\textbf{r},\textbf{v}) \,d^3v\notag\\ 
    = -\frac{e}{\epsilon_0}[n_0-\iiint f_e(\textbf{r},\textbf{v}) \,d^3v], 
\end{gather}
where we have assumed a uniform background ion density, $n_0$. Similarly, we utilize the first-order moment of the distribution function to rewrite Amp\`ere's law as
\begin{equation}
     \nabla\times\textbf{B}=-\mu_0e\iiint f_e(\textbf{r},\textbf{v})\textbf{v} \,d^3v
\end{equation}
where ions are assumed to be stationary and do not contribute to the current density.

Following the recent work of McClung et al., we assume cylindrical symmetry of the electron distribution function, $f_e(\textbf{r},\textbf{v})=f_e(\rho,v_\rho,v_\phi,v_x)$, the electric scalar potential, $\Psi = \Psi(\rho)$, and the magnetic vector potential, $\textbf{A} = A_x(\rho)\hat{x} + A_{\phi}(\rho)\hat{\phi}$, such that $\textbf{B} = \nabla\times\textbf{A} = -\frac{dA_x}{d\rho}\hat{\phi} + \frac{1}{\rho}\frac{d}{d\rho}({\rho}A_{\phi})\hat{x}$. Note that we have chosen $\hat{x}$ to be the axial direction instead of the typical $\hat{z}$  to match the convention of PSC. For uniform \textbf{B}, we require that $A_\phi=B_0\rho/2$, where $B_0$ is the background magnetic field parameter, and $A_x$ be a constant. This uniform \textbf{B} was what we used in previous simulations, but we can now initialize with a nonuniform \textbf{B} that satisfies Amp\`ere's law. 

For normalization, we use the same scheme given in McClung et al.\cite{10.1063/5.0187853} We normalize such that the electron thermal velocity, $v_e$, is the unit for \textbf{v}, the Debye length, $\lambda_D=v_e/\omega_{pe}$, is the unit for \textbf{r}, $n_0e\lambda_D^2/\epsilon_0$ is the unit for $\Psi$, $n_0e\lambda_D/\epsilon_0v_e$ is the unit for \textbf{B}, and $n_0/v_e^3$ is the unit for $f_e$. In the simulations, several physical constants are set to the following:
\begin{subequations}
    \label{eq:normalized 1}
    \begin{align}
        e\ideq \charge{i} = -\charge{e}                                     & \to 1                     \\
        \mass{e}                                                            & \to 1                     \\
        \density{0}=\density{i}=\density{e}\rvert_{\rho=\infty}                         & \to 1                     \\
        \epsilon_0                                                          & \to 1                     \\
        \angfreq{pe} \ideq \sqrt{\frac{e^2\density{e}}{\epsilon_0\mass{e}}} & \to 1 \label{eq:normt}    \\
        \velocity{e}\ideq \sqrt{\frac{\kb\temperature{e}}{\mass{e}}}        & \to 1 \label{eq:normv}  .
    \end{align}
\end{subequations}

With these assumptions and normalizations, the forms of the Vlasov, Poisson, and Amp\`ere equations in cylindrical coordinates respectively become, for an electron distribution function,
\begin{align}
     v_\rho\pderiv{f_e}\rho  + \left(\pderiv{\Psi}{\rho} - v_{\phi}\frac{d({\rho}A_{\phi})}{{\rho}d\rho} -v_x\frac{d(A_x)}{d\rho} + \frac{v_{\phi}^2}{\rho}\right)\pderiv{f_e}{v_{\rho}} \notag\\
     - \left(\frac{v_{\rho}v_{\phi}}{\rho} - v_{\rho}\frac{d({\rho}A_{\phi})}{{\rho}d{\rho}}\right)\pderiv{f_e}{v_{\phi}} 
     + v_{\rho}\frac{d(A_x)}{d{\rho}}\pderiv{f_e}{v_x} = 0, \label{eq:vlasov}
\end{align}

\begin{align}
    \frac{1}{\rho}\pderiv{}{\rho} \left(\rho\pderiv\Psi\rho\right) = \iiint f_e(\rho,v_\rho,v_\phi,v_x)\,d^3v - 1,\label{eq:poisson}
\end{align}
and
\begin{align}
    \frac{d}{d\rho}[\frac{1}{\rho}\frac{d}{d\rho}({\rho}A_\phi)]\hat{\phi}+\frac{1}{\rho}\frac{d}{d\rho}[\rho\frac{dA_x}{d\rho}]\hat{x} \notag\\
    =\beta_{e}^2\iiint f_e(\rho,v_\rho,v_\phi,v_x)\textbf{v}\,d^3v,
\end{align}
where $\beta_e=v_e/c$ and $c$ is the speed of light.

The BGK solution is then an electron distribution function that satisfies these equations self-consistently. Since it has been shown that a distribution function that depends only on the total normalized energy, $w = v^2/2 - \Psi$, is not a valid solution because it does not remain localized,\cite{PhysRevLett.95.245004} we utilize a distribution function that depends on additional constants of motion, such as the normalized canonical angular momentum, $l = \rho(v_\phi - A_\phi)$, and linear momentum, $p = v_x - A_x$. One possible solution is 
\begin{equation}
    f_e(w,l,p)=(2\pi)^{-3/2}e^{-w}[1-h_0\exp(-kl^2-{\xi}p^2)],\label{equation:bgksolution}
\end{equation}
where $h_0$, $k$, and $\xi$ are parameters.\cite{doi:10.1063/1.5126705} In the simulations presented here, $\xi$ is set to 0 for simplicity, so that the solution only depends on electron angular momentum and total energy. The $h_0$ parameter can be set to distinguish between two cases regarding the density of electrons in the center of the structure, which is in the center of the simulation domain. For the EDH, $h_0$ is positive and less than one, and the central normalized electron density is less than unity. For the EDB, $h_0$ is negative and the central normalized electron density is greater than unity. Once this form of the distribution function is assumed and the parameters are assigned, numerical integration is performed to solve for the corresponding electric scalar and magnetic vector potentials to be initialized in the simulation. We also initialize the particle velocities according to the exact method of initialization mentioned in McClung et al. \cite{10.1063/5.0187853}

Because of the form of the distribution function, we can expect instabilities to form. As we showed in our past work with the EDH, the reduced distribution function for the azimuthal velocity, $f_e(v_\phi)$, has a ``bump-on-tail" shape at some radial positions. Velocity distributions of this form are well known to possibly lead to an instability.\cite{chenbook} The distribution function gives a clockwise flow to electrons around the center. Since the bump on the tail is at a positive value of $v_\phi$, we expect an instability that propagates in the counter-clockwise direction.
\begin{figure}
    \centering
    \includegraphics[width=\columnwidth]{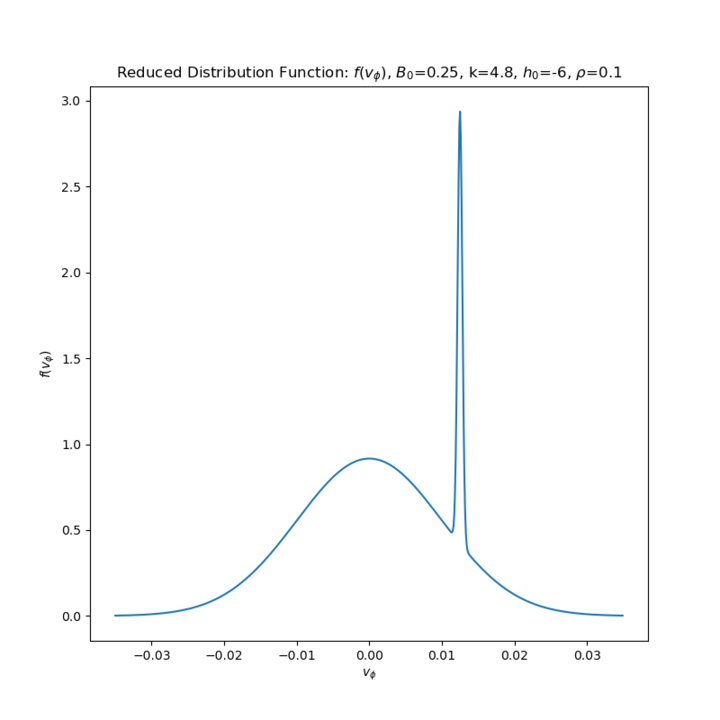}
    \caption{Reduced distribution function corresponding to Eq. (\ref{equation:bgksolution}) for the unstable EDB at $\rho=10\lambda_D$ and $B_0=0.25$.}
    \label{fig:instability}
\end{figure}

\begin{figure}
    \centering
    \includegraphics[width=\columnwidth]{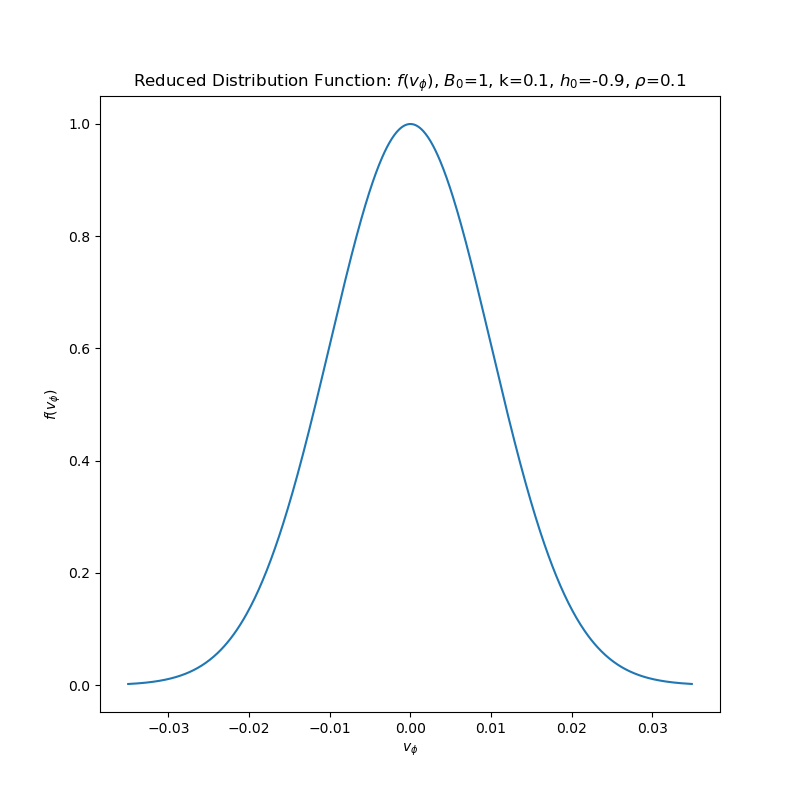}
    \caption{Reduced distribution function corresponding to Eq. (\ref{equation:bgksolution}) for the quasi-stable EDB at $\rho=10\lambda_D$ and $B_0=1$.}
    \label{fig:reduceddiststable}
\end{figure}

Similarly, we can look at the reduced distribution function for $v_\phi$ for the EDB case as well to predict instability. In Figure \ref{fig:instability}, we see the same sort of bump-on-tail shape of the distribution for the lower value of $B_0$. The bump is not as obvious for smaller $\rho$, but becomes more apparent as $\rho$ increases in the distribution function. At $\rho$=10$\lambda_D$, the bump (or spike, rather) is much narrower and sharper than it was for the EDH, leading to unstable behavior. On the other hand, the reduced distribution function of $v_\phi$ is more indicative of a single-humped distribution for larger $B_0$, as shown in Figure \ref{fig:reduceddiststable}. This leads to more stable behavior in the simulations for larger $B_0$.

\subsection{\label{sec:Approximations}Approximations}
We initially ran our simulations without solving Amp\`ere's law for the solution. This was a valid approximation because we set the electron thermal velocity to be a reasonably small fraction of the speed of light. We do this by adjusting the parameter of $\beta_e$. This allowed us to approximate the normalized $\nabla\times\textbf{B}=-\beta_e^2\iiint f\textbf{v}\,d\textbf{v} \approx 0$. It was found that the magnetic field variance was minimal for these parameters. See McClung et al. \cite{10.1063/5.0187853} for a more thorough description of this approximation. As mentioned earlier, we have now included Amp\`ere's law in the solution, so we are no longer limited to very small $\beta_e$, as long as it is not too close to unity to invalidate the non-relativistic theory. Since the PSC simulations are relativistic, the fact that there are good agreements between simulations and theory shows that the non-relativistic solutions are valid for the $\beta_e$ values used in this study.

As mentioned previously, the ions are assumed to form a uniform static background. This approximation can be generalized in the future to more realistic distributions of ions, such as a Boltzmann distribution.

\section{\label{sec:SIMULATION setup}Simulation Setup\protect}
\subsection{Normalization and Units \label{sec:psc normalization}}

The setup and units of the simulations are the same as they were in McClung et al. \cite{10.1063/5.0187853} Here are the key points that affect the units of the simulations' spatial and temporal domains.

The normalizations given in
  \cref{eq:normt,eq:normv} imply that the Debye
  length $\debye\ideq \velocity{e}/\angfreq{pe} =
      1$.
The magnetic
  field strength parameter $B_0$ has units
  $e\mass{e}/\density{e}\angfreq{pe}$.
Note that $B_0=\angfreq{ce}/\angfreq{pe}$, where
  $\angfreq{ce}\ideq eB/\mass{e}$ is the electron
  gyrofrequency.
Thus, the electron gyroperiod $\period{ce}\ideq
      2\pi/\angfreq{ce}$ is $2\pi/B_0$.

PSC simulation results and figures shown are normalized according to \cref{eq:normalized 1} with the change
\begin{align}
    c & \to 1.
    \label{eq:normalized 2}
\end{align}
Thus, time still has units of
  $\angfreq{pe}^{-1}$, but distance now has units
  of the electron skin depth, $c\angfreq{pe}^{-1}$.
This multiplies the length scale compared to
  \cref{eq:normalized 1} by a factor of
  $\relvelocity{e}$, or $\lambda_D=\beta_e.$

\subsection{Case Definitions \label{sec:cases}}

There are two main sets of parameters used for the presented simulations. The first set of parameters, labeled as the EDH case, is $h_0 = 0.9$, $k = 0.4$, which is the same set of parameters used in the 2D runs of McClung et al. \cite{10.1063/5.0187853}  Furthermore, within the two cases, we identify two more cases for different values of $B_0$. $B_0=1$ will be the ``quasi-stable" case, and $B_0=0.25$ will be the ``unstable case." This identification follows from our findings in McClung et al. that $B_0=0.25$ produced an instability quicker than $B_0=1$. \cite{10.1063/5.0187853} 
The EDB uses two other sets of parameters. This was done because it was found that a similar set of parameters to the EDH produced much more stable results. Therefore, for the EDB quasi-stable case, $h_0=-0.9$, and $k=0.4$ were used. For the unstable case, $h_0=-6.0$, and $k=4.8$ were used; this second set of parameters produced a qualitatively similar initial electric potential to the unstable EDH run. An interesting feature of a EDB solution is that it has a negative electric potential so that there is no electrostatic electron trapping. This is not possible for a localized 1D BGK mode, since the electron trapping is essential for the existence of solutions.

Within these cases, we include runs with the generalizations we made to the solution and simulation. To distinguish the cases where we change the $\beta_e$ parameter, we will explicitly list the $\beta_e$ parameter being used in the simulation. We will also list whether the simulation used VP initialization or VM initialization. When addressing the individual simulations, we will use shorthand to denote the run being discussed (e.g., EDH-VP represents a run that is an electron density hole that was initialized using the Vlasov-Poisson method of initialization). A table of cases being discussed in this paper (along with resolutions) is included in Table \ref{runslist}.

\begin{table}[]
    \centering
    \begin{tabular}{ | m{0.75cm} | m{5em} | m{1.5cm}| m{1cm} | m{1.8cm} |} 
  \hline
  $B_0$ & EDH, EDB & VP, VM & $\beta_e$  & Resolution\\ 
  \hline
  1  & EDH & VP & 0.001 & $512^2$ (2D) \\ 
  \hline
  1  & EDH & VP & 0.001 & $256^3$ (3D) \\ 
  \hline
    1  & EDB & VM & 0.01 & $1024^2$ (2D) \\ 
  \hline
      1  & EDB & VM & 0.01 & $256^3$ (3D) \\ 
  \hline
  0.25  & EDH & VP & 0.001 & $1024^2$ (2D) \\ 
  \hline
      0.25  & EDH & VP & 0.001 & $256^3$ (3D) \\ 
  \hline
      0.25  & EDB & VP & 0.001 & $256^2$ (2D)\\ 
  \hline
      0.25  & EDB & VP & 0.01 & $256^2$ (2D)\\ 
  \hline
      0.25  & EDB & VM & 0.01 & $1024^2$ (2D) \\ 
  \hline
      0.25  & EDB & VM & 0.01 & $256^3$ (3D)\\ 
  \hline
\end{tabular}
    \caption{Runs being considered. Note that all 3D runs have the same resolution.}
    \label{runslist}
\end{table}

\subsection{Initialization \label{sec:parameters}}
The same method of velocity initialization outlined in our previous work is used in these simulations as well. Initialized fields and quantities, such as $\Psi$, $\textbf{B}$, charge density $\rho_q$,  are initialized in the same way as McClung, et al. \cite{10.1063/5.0187853} for the VP cases, such that they were evaluated at discreet values of $\rho$ and interpolated onto the simulation grid. In 3D, each level of the simulation in the axial direction is initialized with different random particle velocities following the same distribution function. For the VM cases, the same methods of initialization are used for all fields and quantities except for \textbf{B}. We now utilize Amp\`ere's law to solve for \textbf{B} and interpolate it onto the grid, instead of the uniform \textbf{B} used in our previous work. Finally, the simulations are all initialized with 100 particles per cell.

\section{SIMULATION RESULTS \& DISCUSSION \label{sec:results&discussion}}

Eight new runs are presented here, listed in Table \ref{runslist}. We will also make use of some of the simulation results from McClung et al. \cite{10.1063/5.0187853} We show a $512^2$ resolution quasi-stable EDH-VP run and a $1024^2$ resolution unstable EDH-VP run. Because of the significantly increased computation time for higher resolution 3D runs, high-resolution runs are done in 2D to show the evolution of the simulation in the perpendicular plane in greater spatial detail. Again, following the PSC convention, the direction along the background magnetic field is in the positive $\hat{x}$ direction. The majority of the visualizations shown here show the structure from a perspective looking anti-parallel to the background magnetic field, showing the perpendicular $y$-$z$ plane. Because of the nature of 3D simulations, it is difficult to present the results in the format of a paper. Therefore, most of the 3D simulation analysis presented here is done by looking at these 2D cross-sections at different levels along the axial direction. We have included videos of our 3D simulations from different perspectives showing more of the structure in the supplemental material. For all of the presented 3D simulations, the simulation domain in the directions parallel and perpendicular to the background magnetic field are equivalent. 

There are several different results we wish to look at from our simulations. First, we want to see if the addition of the third dimension adds any new dynamics to the 2D case we studied in our previous paper and discuss these new results. We then want to see if the EDB has similar instability behavior to the EDH, both in 2D and 3D simulations. Lastly, we will look at the impacts on the simulation from our new generalizations.

\subsection{Electron Density Hole (EDH)\label{sec:EDH}}
For the EDH, we ran 2D and 3D simulations for two different values of $B_0$ that were found to be either quasi-stable or unstable from our previous work in 2D. All of the presented EDH runs use $\beta_e=0.001$ unless otherwise noted.

\begin{figure}
    \centering
    \includegraphics[width=\columnwidth]{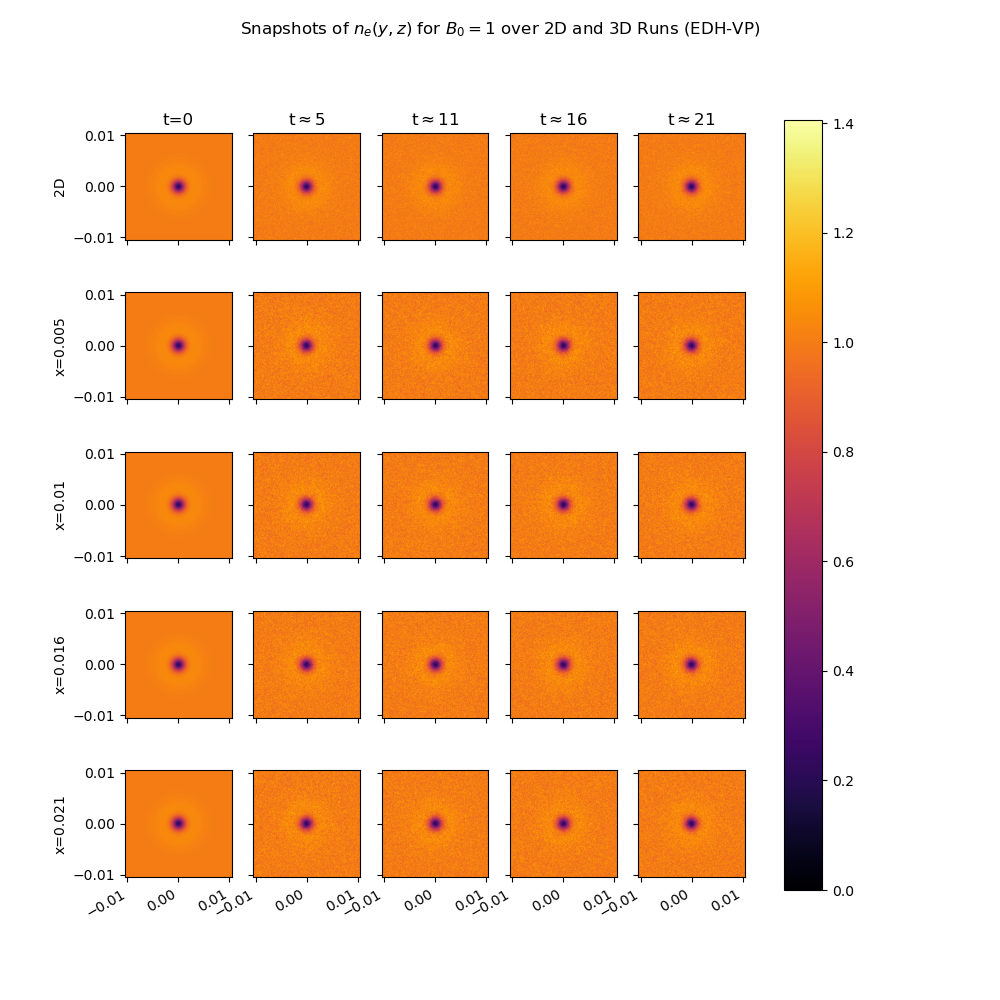}
    \caption{Quasi-stable EDH density snapshots. $x$ values correspond to 1/4, 1/2, 3/4, and 1 times the length of the simulation in the third dimension.}
    \label{fig:stableEDHdensitysnapshots}
\end{figure}

The spatial domain volume for the quasi-stable 3D EDH-VP run was around $8,000\lambda_D^3$ or (20$\lambda_D)^3$, which was determined following the same scheme as McClung et al. for the 2D simulations. \cite{10.1063/5.0187853} In this case, we found that the 3D results were qualitatively similar to the 2D results. The instability that was seen in our previous work in 2D takes a long time (around $t\approx400$) to develop. Because of the length of time required to run the simulation until the onset of the instability, we limited the 3D run to a just a few electron gyroperiods to see if the initial stable equilibrium was altered by the inclusion of the third dimension. Notably, there seems to be no new instability that develops because of the added third dimension. Essentially, the 3D quasi-stable EDH-VP simulation seems to confirm the time-steadiness, at least initially, in the same way as the 2D run of the same name. This can be seen more clearly in Figure \ref{fig:stableEDHdensitysnapshots}.

\begin{figure}
    \centering
    \includegraphics[width=\columnwidth]{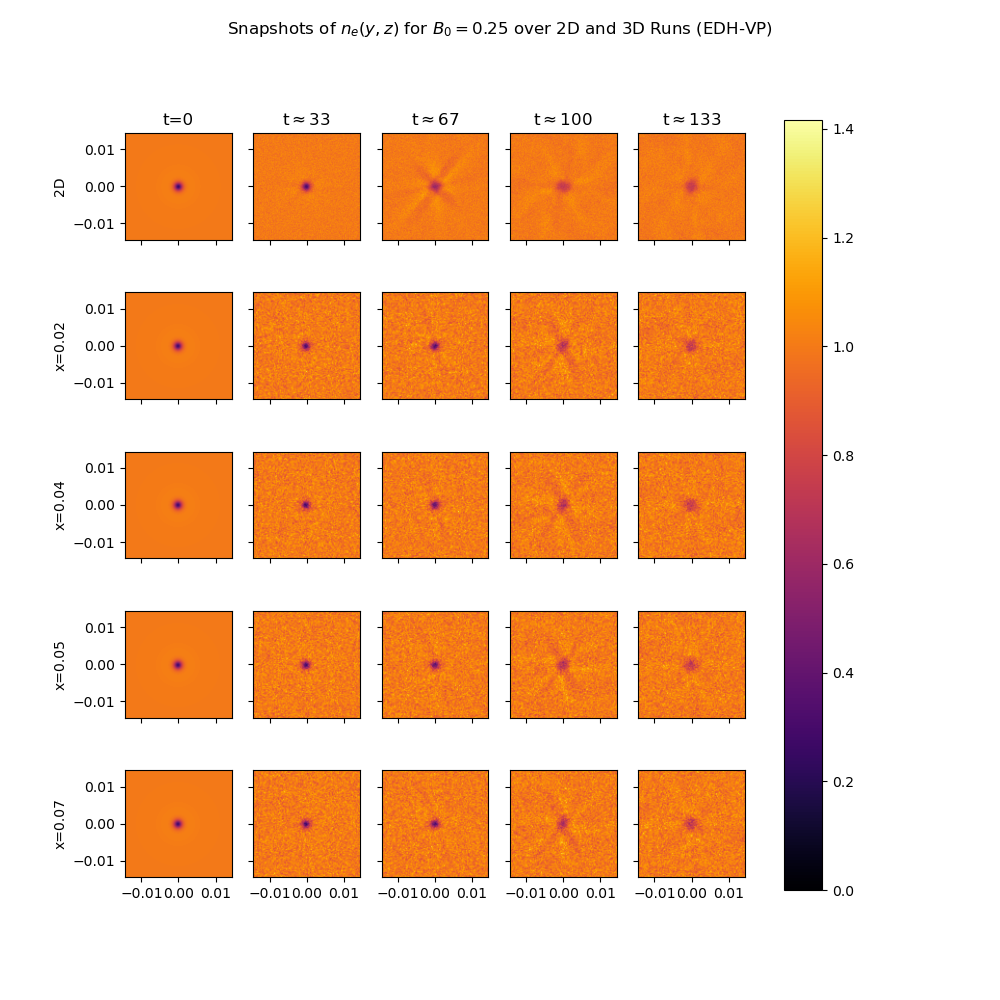}
    \caption{Unstable EDH density snapshots. $x$ values correspond to 1/4, 1/2, 3/4, and 1 times the length of the simulation in the third dimension.}
    \label{fig:unstableEDHdensitysnapshots}
\end{figure}

For the unstable 3D EDH-VP simulation, we utilized a simulation spatial domain size of around 70$\lambda_D$ in each direction, giving a spatial resolution that is 0.27$\lambda_D$. This simulation domain size is again decided according to the convention outlined in McClung et al. In this case, we saw different results between the 2D and 3D runs, as can be seen in Figure \ref{fig:unstableEDHdensitysnapshots}.

\begin{figure}
    \centering
    \includegraphics[width=\columnwidth]{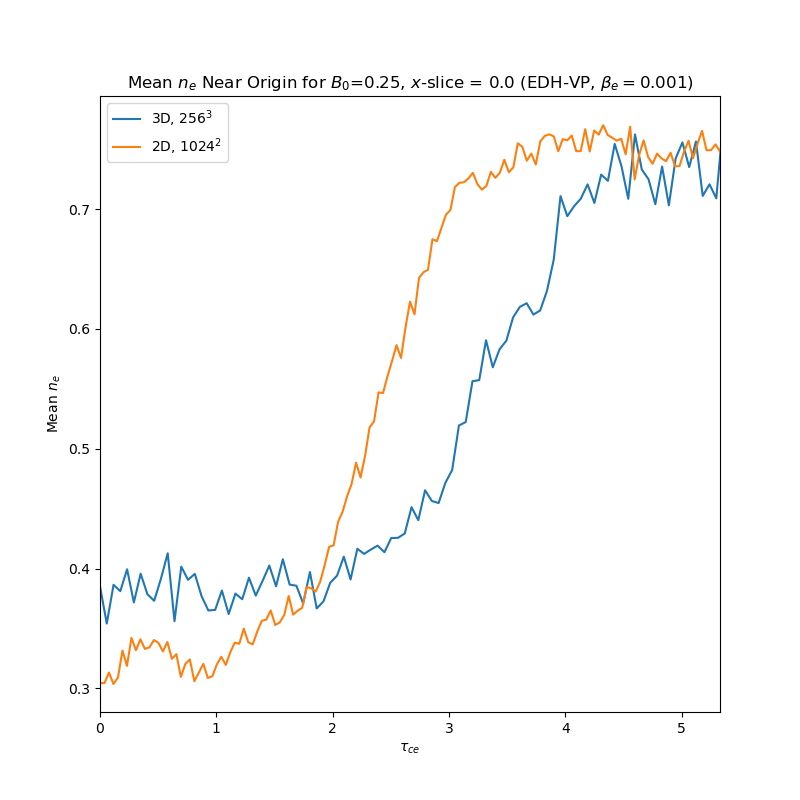}
    \caption{Mean electron density at the origin comparison between 2D and 3D unstable EDH-VP runs}
    \label{fig:EDHdelay}
\end{figure}

Qualitatively, the two runs are identical at the start. However, the 3D run seems to maintain stability for a longer time than the 2D run. We can see from the mean value of $n_e$ at the center of the simulation that the 3D run takes $\approx1\tau_{ce}$ longer to reach a similar level of electron density hole filling to the 2D case, which can be seen in Figure \ref{fig:EDHdelay}. Like in our previous work, this value is calculated as the average $n_e$ across the middle four grid cells for runs with a resolution of $256^2$ or an equivalent area for higher resolution runs.\cite{10.1063/5.0187853}

\begin{figure}
    \centering
    \includegraphics[width=\columnwidth]{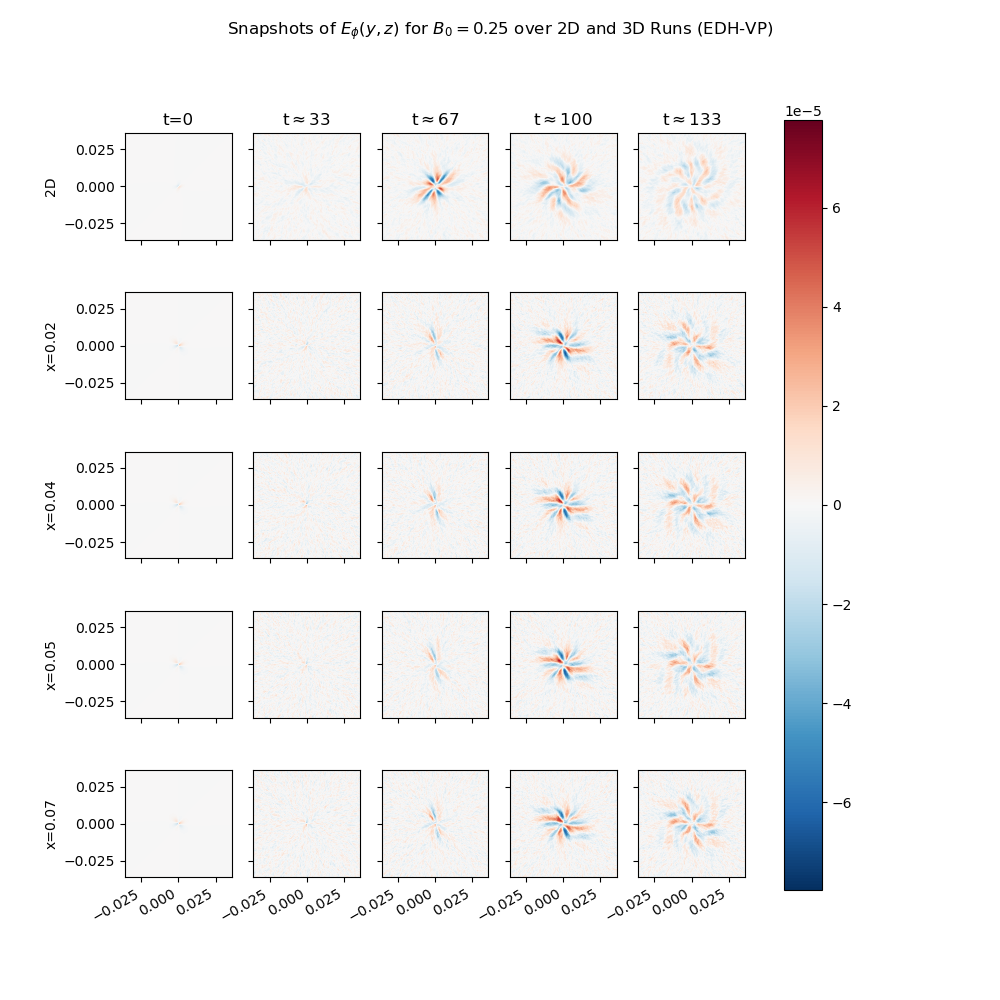}
    \caption{Unstable EDH azimuthal electric field snapshots. $x$ values correspond to 1/4, 1/2, 3/4, and 1 times the length of the simulation in the third dimension.}
    \label{fig:unstableEDHephisnapshots}
\end{figure}

\begin{figure}
\centering
    \includegraphics[width=\columnwidth]{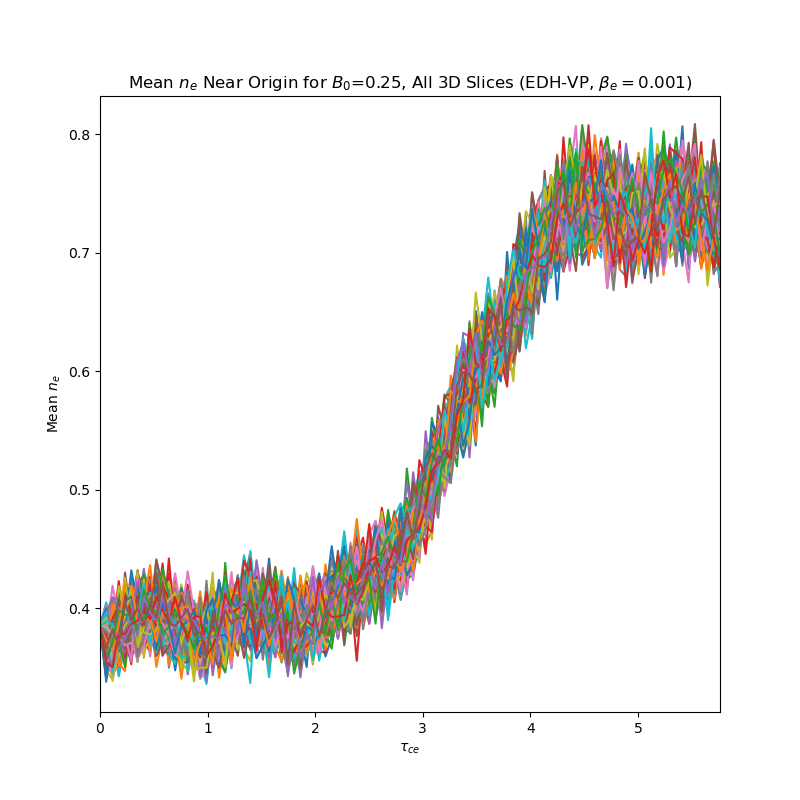}
    \caption{Mean electron density at the center of the simulation at different levels.}
    \label{fig:EDH3dneallslices}
\end{figure}

Once both of the runs reach a similar time in their instabilities, the electrostatic wave that propagates in the azimuthal direction that we saw in our 2D runs becomes visible, as can be seen in Figure \ref{fig:unstableEDHephisnapshots}. 
One remarkable feature of the 3D run is that it seems that the instability is in-phase between all levels of the simulation. This in-phase behavior is apparent when comparing the almost identical bottom four rows of Figures \ref{fig:unstableEDHdensitysnapshots} and \ref{fig:unstableEDHephisnapshots}. It seems that the instability is slightly out of phase initially, but becomes in-phase over a short time scale. As a consequence of the structure evolving in-phase on all levels of the simulation, we see that there is only a slight deviation of the average $n_e$ at the center of the simulation from level to level, depicted in Figure \ref{fig:EDH3dneallslices}. It is possible that Landau damping is the cause for this in-phase nature of the simulation; however, more work needs to be done to explain this phenomenon.

\subsection{Electron Density Bump (EDB) \label{sec:EDB}}

Next, we present the results of the EDB-VM runs with $\beta_e=0.01$ unless otherwise specified. This means that the spatial domain size being used in these simulations will be a factor of 10 larger than the domain sizes used in the EDH case. We will defer the discussion of the validity of the VM initialization with a larger $\beta_e$ in Sub-section C below. For now, we will only look at the results of the quasi-stable and unstable EDB simulations. These cases have a few similarities to the EDH cases, but there are differences.

\begin{figure}
    \centering
    \includegraphics[width=\columnwidth]{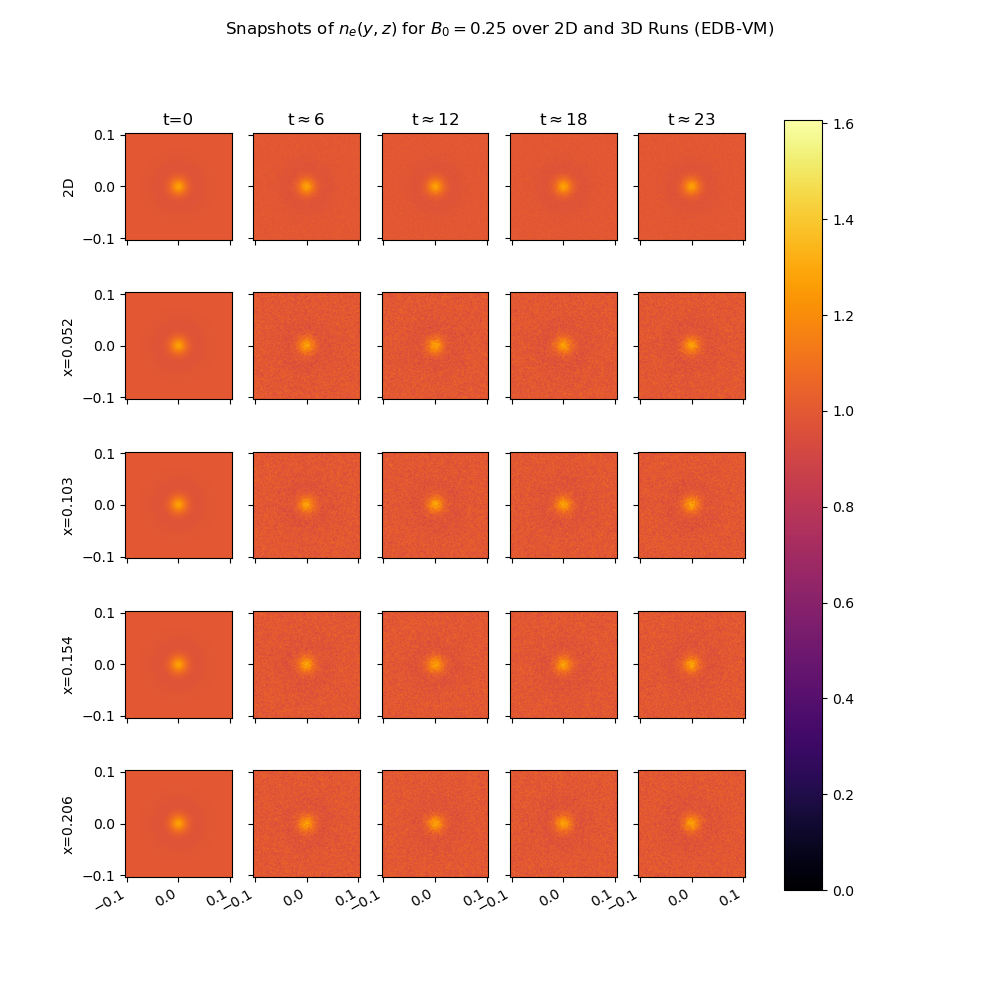}
    \caption{Quasi-stable EDB density snapshots. $x$ values correspond to 1/4, 1/2, 3/4, and 1 times the length of the simulation in the third dimension.}
    \label{fig:edbstabledensitysnapshots}
\end{figure}

For the quasi-stable case, we see the same behavior in both 2D and 3D, much like the EDH. A comparison of the quasi-stable case is shown in Figure \ref{fig:edbstabledensitysnapshots}. Similar to the EDH, we found that an instability starts to develop in 2D at around $t\approx400$, and the added third dimension does not induce an additional instability.

\begin{figure}
    \centering
    \includegraphics[width=1\columnwidth]{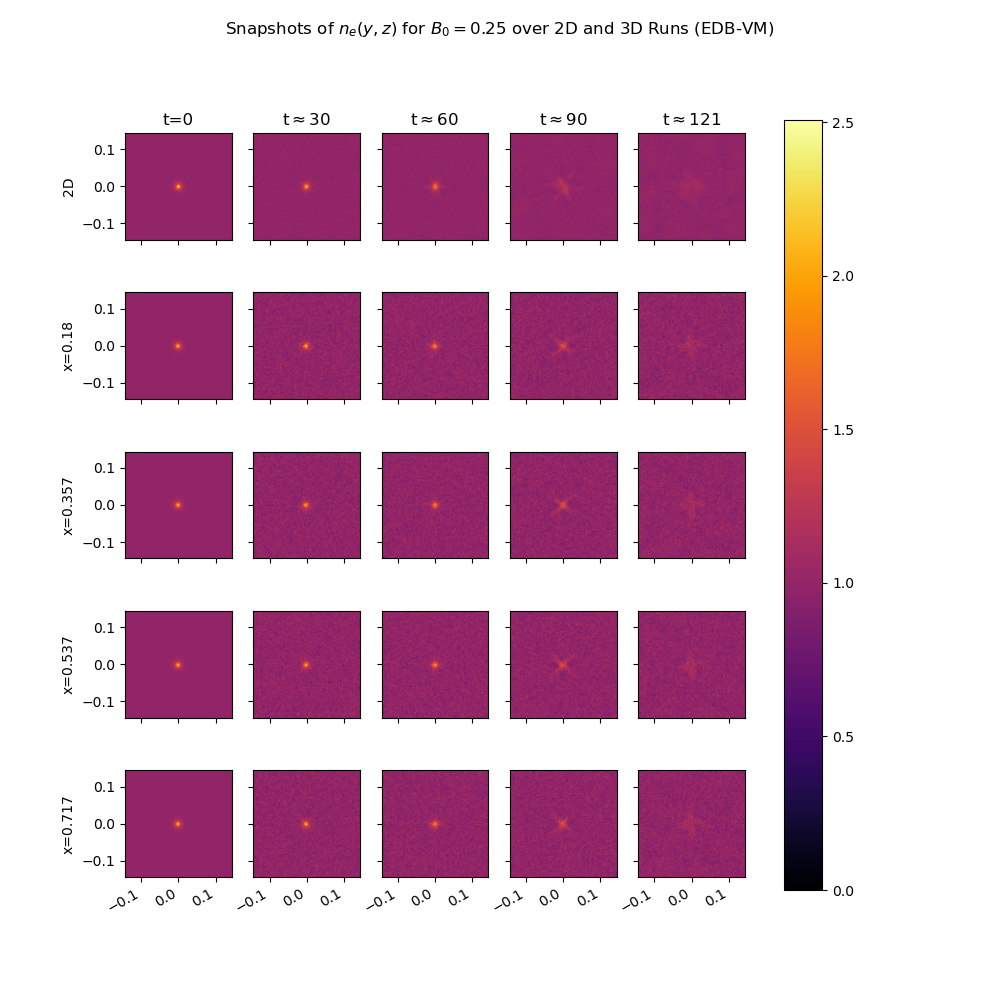}
    \caption{Unstable EDB density snapshots. $x$ values correspond to 1/4, 1/2, 3/4, and 1 times the length of the simulation in the third dimension.}
    \label{fig:edbunstabledensitysnapsots}
\end{figure}

\begin{figure}
    \centering
    \includegraphics[width=\columnwidth]{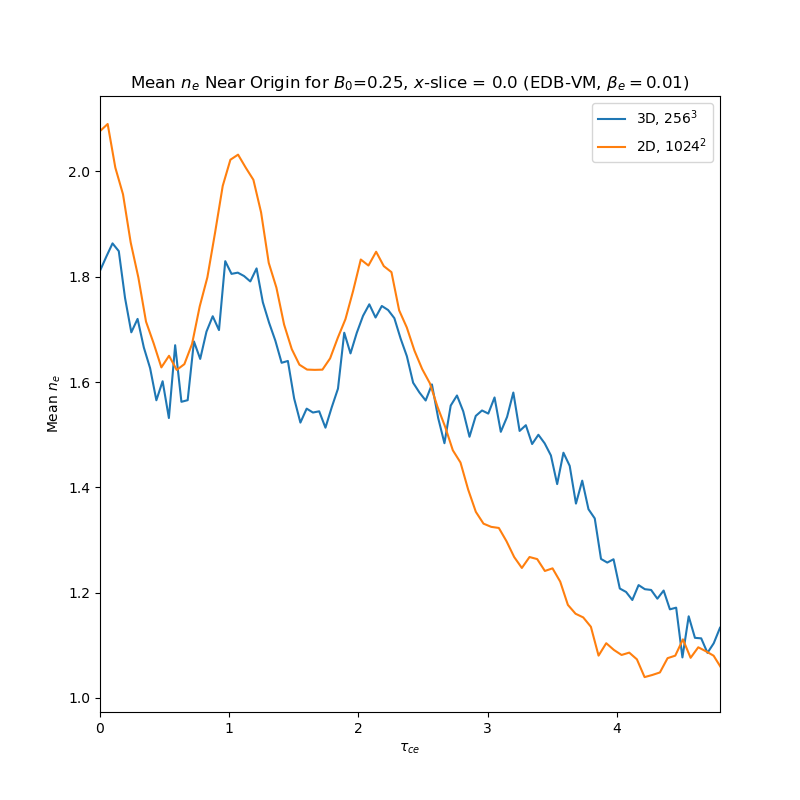}
    \caption{Mean $n_e$ near the origin for the 2D and 3D unstable EDB runs.}
    \label{fig:EDBdelay}
\end{figure}

\begin{figure}
\centering
    \includegraphics[width=\columnwidth]{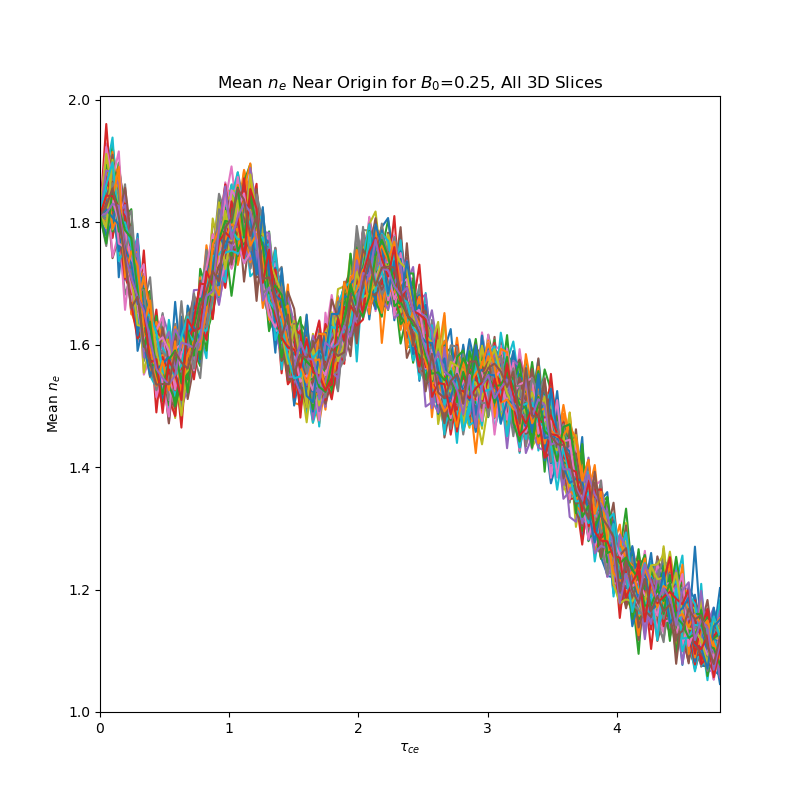}
    \caption{Mean $n_e$ near the origin at each level of the 3D unstable EDB run.}
    \label{fig:EDB3dneallslices}
\end{figure}

The much more notable runs for the EDB-VM case can be seen in the unstable case. Here we see some similarities to the EDH, but also some key differences. First, we see in Figure \ref{fig:edbunstabledensitysnapsots} that the instability again takes longer to develop in 3D than it does in 2D. Also, we see that there is a significant oscillation of the density bump in both 2D and 3D before the instability develops. Figure \ref{fig:EDBdelay} shows the delay in instability development as well as the oscillations, and Figure \ref{fig:EDB3dneallslices} shows the electron density oscillations. This oscillation seems to occur twice over the course of 2$\tau_{ce}$ with a smaller third oscillation in 3D at around 3$\tau_{ce}$. Again, the delay in 3D seems to be about 1$\tau_{ce}$.

\begin{figure}
    \centering
    \includegraphics[width=1\columnwidth]{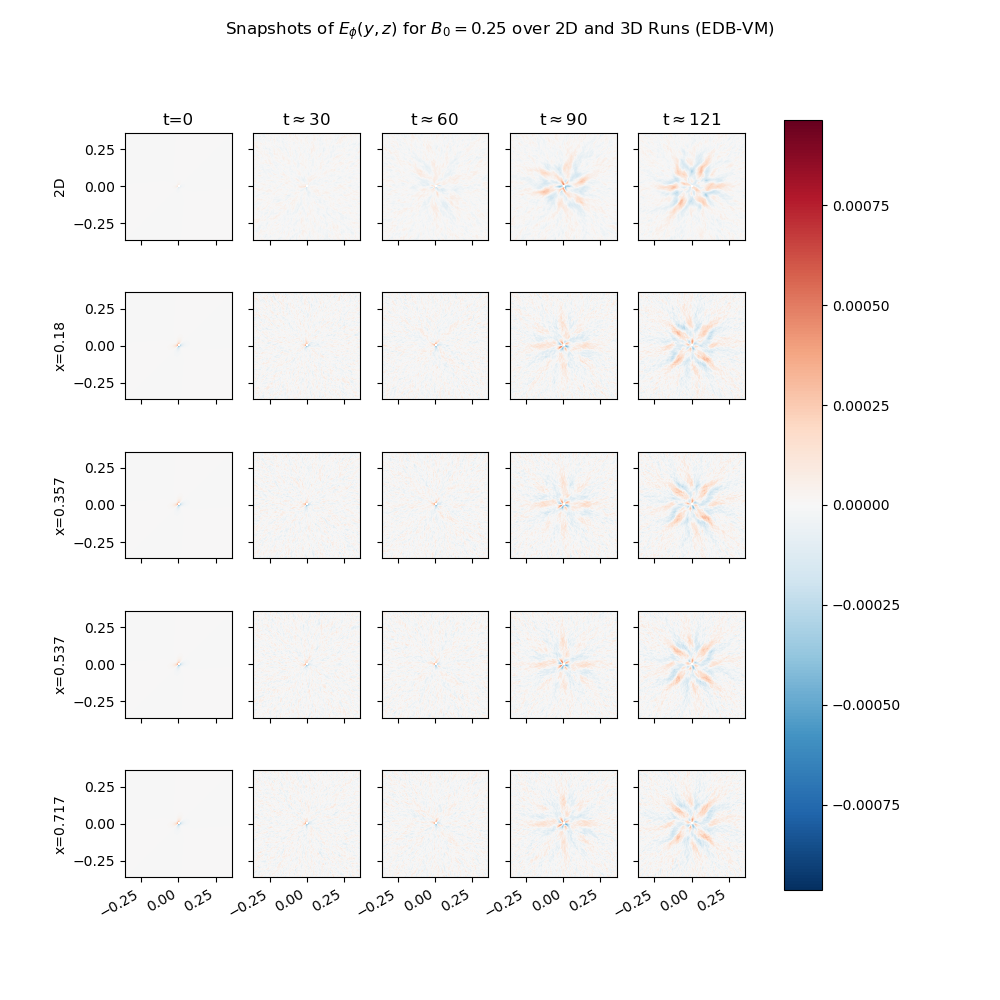}
    \caption{Unstable EDB azimuthal electric field snapshots. $x$ values correspond to 1/4, 1/2, 3/4, and 1 times the length of the simulation in the third dimension.}
    \label{fig:edbdephisnapsots}
\end{figure}

Additionally, we see the same in-phase behavior we saw in the EDH case. The azimuthal electric field heat map comparisons shown in Figure \ref{fig:edbdephisnapsots} again show in-phase behavior in sharper detail. One other feature of this case is the existence of what seems to be two concentric spirals: one smaller amplitude spiral in the center, and a larger amplitude spiral towards the edge of the simulation (however, these spirals could just as well be parts of one larger spiral with a sharp discontinuity at $\rho\approx0.1$). As the simulation continues, the center spiral seems to fade away, leaving only the larger spiral on the outside. It is currently unknown why these two spirals form in this case and not in the EDH case, and more work is required to understand them. 

\subsection{Discussion of 3D Dynamics\label{sec:delayandinphase}}

In this section, we present two possible reasons for the observed delay of the instability in 3D. We also further discuss the in-phase characteristic of the 3D runs.  

The first possible reason for the delay is a reduction in particle noise in the simulation. We saw in our previous work that a decrease in particle noise (an increase in particles-per-cell and, therefore, total particles) leads to a delay in 
 the start of the instability for runs that were otherwise the same. \cite{10.1063/5.0187853}  Since the 2D and 3D runs presented here all have the same number of particles-per-cell, the total number of particles is greater in the 3D runs and the particle noise is therefore reduced, so we see this effect. 
 
 The second reason for the delay is attributed to the in-phase nature of the instability, the other notable feature of the 3D runs. It is interesting that it seems all levels of the simulation ``wait" to work out the fastest growing mode and become unstable afterwards. This ``waiting" could also be characterized as fluctuations of the electric field and electron density along the axial direction becoming Landau damped. If there were significant changes in electron density between levels of the simulation, the axial electric field, $E_x$, would become large. Landau damping in the simulation keeps this from occurring. This leads to the same instability spiral pattern on all levels developing for the 3D simulation, instead of many different unstable modes evolving at different levels. This period of delay between all levels is what contributes  to the overall delay of the instability. It is the combined effects of reduced particle noise and Landau damping that leads to a delay in the instability in 3D, as well as the instability evolving in-phase.

\subsection{Impacts of Generalization \& Future Work \label{sec:generalizationsfuturework}}

Here, we look at the two generalizations we have made to the assumed solution and simulations. Namely, we wish to investigate the use of larger $\beta_e$ and VM initialization. If we can make these generalizations to the BGK mode solutions and simulations, it will have positive effects moving forward.

As for the larger $\beta_e$, it was found that there was no notable difference between the results from two runs that only differed in their values of $\beta_e$, aside from a change in spatial scale and velocity scale. For the sake of space and having fewer figures here, we simply acknowledge this is the case and leave these two sets of results in the supplementary material. Since we have identified that this is the case, we are no longer limited to small values of $\beta_e$. This means we can use values of $v_e$ that are better suited for our approximation of a collisionless plasma. Also, because of the nature of the Courant–Friedrichs–Lewy (CFL) condition that PSC uses, since these higher $\beta_e$ cases result in a simulation that has a larger spatial scale, then they also have a larger time step. \cite{GERMASCHEWSKI2016305} Therefore, as we increase $\beta_e$, the computation time required to run a simulation to a desired physical time decreases. Also, a larger $\beta_e$ means that we can no longer assume a constant background magnetic field, and Amp\`ere's law must be solved.

\begin{figure}
\begin{subfigure}{\columnwidth}
    \centering
    \includegraphics[width=\columnwidth]{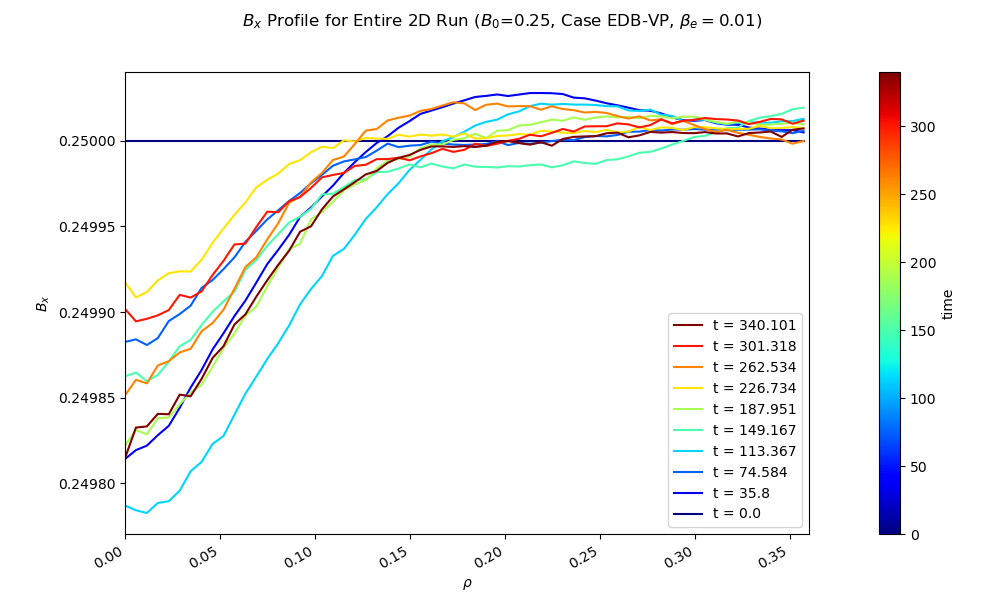}
    %\caption{Vlasov-Poisson initialization}
    \label{fig:BxVP}
\end{subfigure}
\begin{subfigure}{\columnwidth}
    \centering
    \includegraphics[width=\columnwidth]{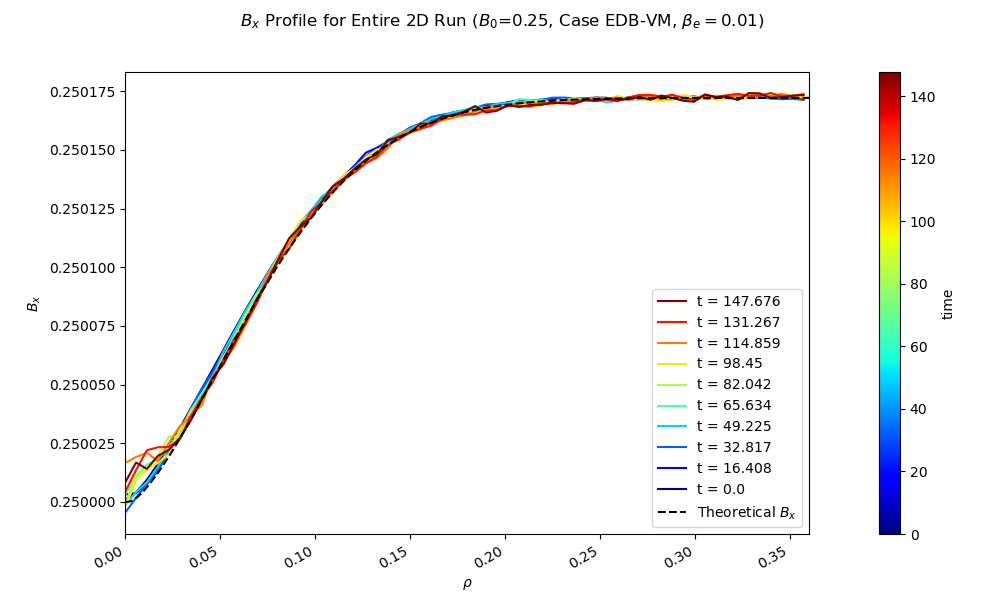}
    %\caption{Vlasov-Maxwell initialization}
    \label{fig:BxVM}
\end{subfigure}
\caption{Plots of $B_x$ profiles showcasing the difference between VP (top) and VM (bottom) initialization over the course of the simulations for $B_0=0.25$ EDB ($256^2$ resolution) The black dashed line in the VM plot represents the initial $B_x$ that was found by solving Amp\`ere's law.}
\label{fig:Bxplots}
\end{figure}

\begin{figure}
\begin{subfigure}{\columnwidth}
    \centering
    \includegraphics[width=\columnwidth]{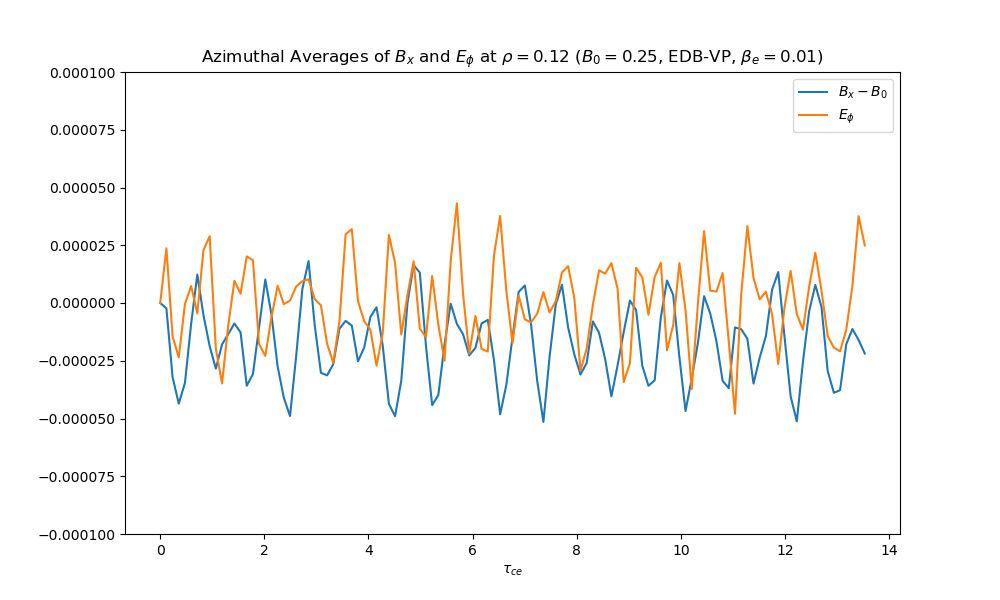}
    %\caption{Vlasov-Poisson initialization}
\end{subfigure}
\begin{subfigure}{\columnwidth}
    \centering
    \includegraphics[width=\columnwidth]{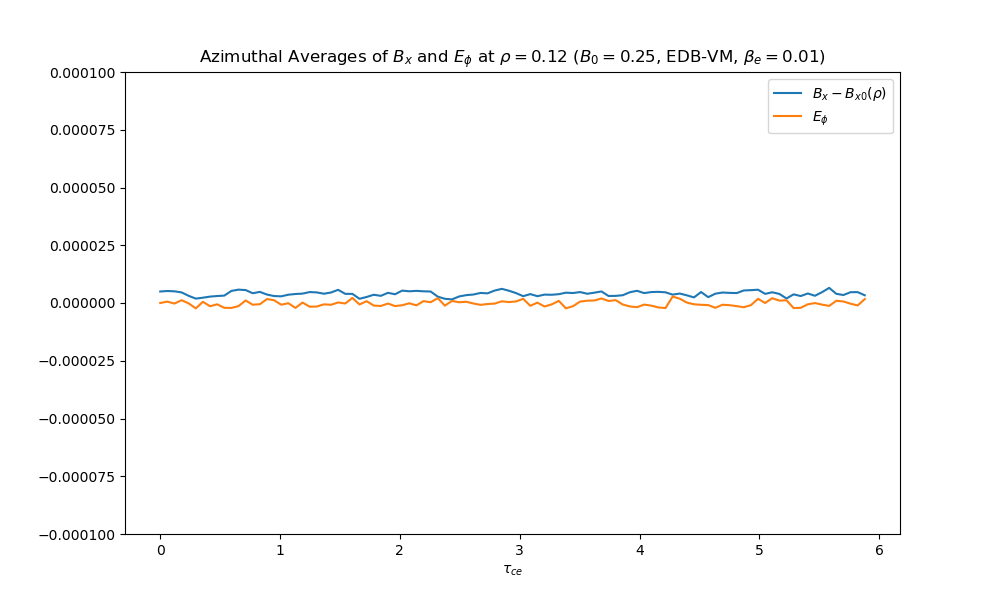}
    %\caption{Vlasov-Maxwell initialization}
\end{subfigure}
\caption{Plots of $B_x$ and $E_\phi$ showcasing the difference between VP (top) and VM (bottom) initialization over the course of the simulations for $B_0=0.25$ EDB ($256^2$ resolution) at $\rho=0.12$. The $B_x$-$B_0$ plotted (blue) is the deviation from the initialized value of $B_0$ in both simulations.}
\label{fig:bxephivariance}
\end{figure}

Our generalization of solving for Amp\`ere's law does seem to have an impact on the time-steadiness of a few field quantities. This is to be expected, since, while the value of $\beta_e$ was small in the VP runs, it was nonzero. This leads to a variation of the initially uniform magnetic field on the order of $\beta_e^2$ from Eq. (10). As can be seen in Figure \ref{fig:Bxplots}, the VP method of initialization has this small variance and oscillates in space and time, while the VM initialization does not. These plots consist of azimuthal averages of $B_x$ as a function of $\rho$ at different time-steps of the simulation. The VM initialization results in an initial equilibrium of $B_x$, which then begins to vary over time as the instability develops towards the end of the simulation. The deviation from the theoretical value of $B_x$ near $\rho\approx0$ is due to this instability, which is the azimuthal electrostatic wave mentioned earlier. Additionally, we saw a decrease in the time variation of $E_\phi$ when utilizing the VM method. In the VP case, the significantly time varying $B_x$ would induce a significant change in $E_\phi$ from Maxwell's equations. For VM, this $B_x$  variation is nearly eliminated at larger $\rho$, so we no longer see a significant change in $E_\phi$. This difference is seen in Figure \ref{fig:bxephivariance}. Moving forward, we plan to use the VM method of initialization, as it is more exact and leads to more time-steady results for $B_x$ and $E_\phi$. The VM method of initialization did not appear to alter the overall instability seen in the VP runs.

More work is needed to explain the in-phase nature of the 3D runs, as well as the concentric spirals seen for the EDB. Currently, a paper discussing the nature of the EDH instability, such as its magnitude and growth rate, is in progress. Also, more visualization that can highlight the differences between the 2D and 3D cases is desired.

In the future, we plan to also investigate how this BGK mode evolves when the $\xi$ parameter is nonzero. More work is needed to be able to initialize this case, so we anticipate this to be in our future work.

\section{CONCLUSIONS \label{sec:conclusions}}

We found that adding a third dimension to our past EDH simulations did not affect the overall dynamics of the structure. Quasi-stable 2D runs were still stable in 3D, meaning there was no new instability resulting from the added dimension. For the unstable cases, the new dimension only delayed the onset of the instability for about one electron gyroperiod. This does not necessarily mean that the 3D runs are more stable than the 2D runs. The 3D runs are just as unstable as the 2D runs and develop into the same unstable mode with spiral structures; however, the 3D run is slightly delayed, and the spiral structure is in-phase along the axial direction. We attribute this delay to reduced particle noise and Landau damping. 

With regard to our generalizations, we found that our second set of parameters, or the EDB, produced qualitatively similar results to the EDH. In particular, quasi-stable runs were still quasi-stable and unstable runs were still unstable; but, the onset of the instability was delayed in 3D and was in-phase on all levels in the axial direction. However, a more complicated spiral structure formed when the EDB became unstable, and the EDB had more apparent oscillations of mean electron density in the center of the structure. Furthermore, we found that increasing the electron thermal velocity of the simulations by a factor of 10 did not have any noticeable effects on the evolution of the instability in either 2D or 3D, and the VM method of initialization results in an equilibrium of the slightly non-uniform $B_x$ at the start of the simulation, which is time-steady up until the instability develops.

\begin{acknowledgments}
    This work is supported by National Science
      Foundation grants PHY-2010617 and PHY-2010393.
    This research used resources of the National
      Energy Research Scientific Computing Center, a
      DOE Office of Science User Facility supported by
      the Office of Science of the U.S.
    Department of Energy under Contract No.
    DE-AC02-05CH11231.
\end{acknowledgments}

\section*{Data Availability}
The data that supports the findings of this study are available within the article and its supplementary material. Additional data are available from the corresponding author upon reasonable request.

\section*{Supplementary Material}

\url{https://drive.google.com/drive/folders/1h4V4FLA8tPDN-FOVabTCYEjYg2Jh-BV_?usp=drive_link}

\section*{REFERENCES}
%\bibliography{paper1-sources}

%merlin.mbs aipnum4-1.bst 2010-07-25 4.21a (PWD, AO, DPC) hacked
%Control: key (0)
%Control: author (8) initials jnrlst
%Control: editor formatted (1) identically to author
%Control: production of article title (-1) disabled
%Control: page (0) single
%Control: year (1) truncated
%Control: production of eprint (0) enabled
\providecommand{\noopsort}[1]{}\providecommand{\singleletter}[1]{#1}%

\end{document}

%% file: preamble.tex
\usepackage{amsmath,amssymb,amsthm}
\usepackage{enumitem}
\usepackage{mathtools}
\usepackage{esvect}
\usepackage[margin=1in]{geometry}
\usepackage{ifthen}

\let\ideq\equiv

\newcommand{\pderiv}[2]{\frac{\partial #1}{\partial #2}}

\newtheorem*{axiom*}{Axiom}

\newtheorem*{defn*}{Definition}

\newtheorem*{theorem*}{Theorem}

\newtheorem*{metatheorem*}{Metatheorem}

\newtheorem*{lemma*}{Lemma}

\newtheorem*{cor*}{Corollary}

\newtheorem*{remark*}{Remark}

\newtheorem*{prop*}{Proposition}